\documentclass[letterpaper]{article} 
\usepackage{aaai2026}  
\usepackage{times}  
\usepackage{helvet}  
\usepackage{courier}  
\usepackage[hyphens]{url}  
\usepackage{graphicx} 
\usepackage{natbib}  
\usepackage{caption} 
\usepackage{algorithm}
\usepackage{algorithmic}
\usepackage[most]{tcolorbox}  
\usepackage{booktabs}
\usepackage{multirow}
\usepackage{transparent}
\usepackage{amsmath}
\usepackage{amssymb}
\usepackage{graphicx}           
\usepackage{booktabs}           
\usepackage{colortbl}
\usepackage{newfloat}
\usepackage{listings}
\DeclareCaptionStyle{ruled}{labelfont=normalfont,labelsep=colon,strut=off} 
\floatstyle{ruled}
\newfloat{listing}{tb}{lst}{}
\floatname{listing}{Listing}
\title{Listening Between the Frames: Bridging Temporal Gaps in Large Audio-Language Models}
\author{
  Hualei Wang\textsuperscript{1,2}\thanks{Equal contribution.}
  Yiming Li\textsuperscript{1,2}\footnotemark[1]\
  Shuo Ma\textsuperscript{1,2}
  Hong Liu\textsuperscript{1}
  Xiangdong Wang\textsuperscript{1}\thanks{Corresponding author.}
}
\affiliations{

    $^1$Beijing Key Laboratory of Mobile Computing and Pervasive Device, \\
    Institute of Computing Technology, Chinese Academy of Sciences, Beijing, China \\
    $^2$University of Chinese Academy of Sciences, Beijing, China \\ 
    {wanghualei23s,liyiming22s,mashuo20g,hliu,xdwang@ict.ac.cn}
}

\usepackage{bibentry}

\begin{document}

\maketitle

\begin{abstract}
Recent Large Audio-Language Models (LALMs) exhibit impressive capabilities in understanding audio content for conversational QA tasks. However, these models struggle to accurately understand timestamps for temporal localization (e.g., Temporal Audio Grounding) and are restricted to short audio perception, leading to constrained capabilities on fine-grained tasks. We identify three key aspects that limit their temporal localization and long audio understanding: (i) timestamp representation, (ii) architecture, and (iii) data. To address this, we introduce TimeAudio, a novel method that empowers LALMs to connect their understanding of audio content with precise temporal perception. Specifically, we incorporate unique temporal markers to improve time-sensitive reasoning and apply an absolute time-aware encoding that explicitly grounds the acoustic features with absolute time information. Moreover, to achieve end-to-end long audio understanding, we introduce a segment-level token merging module to substantially reduce audio token redundancy and enhance the efficiency of information extraction. Due to the lack of suitable datasets and evaluation metrics, we consolidate existing audio datasets into a new dataset focused on temporal tasks and establish a series of metrics to evaluate the fine-grained performance. Evaluations show strong performance across a variety of fine-grained tasks, such as dense captioning, temporal grounding, and timeline speech summarization, demonstrating TimeAudio's robust temporal localization and reasoning capabilities. 

\textbf{Code} --- \url{https://github.com/lysanderism/TimeAudio}

\end{abstract}
%

\section{Introduction}
\begin{figure}
    \centering
    \includegraphics[width=1.\linewidth]{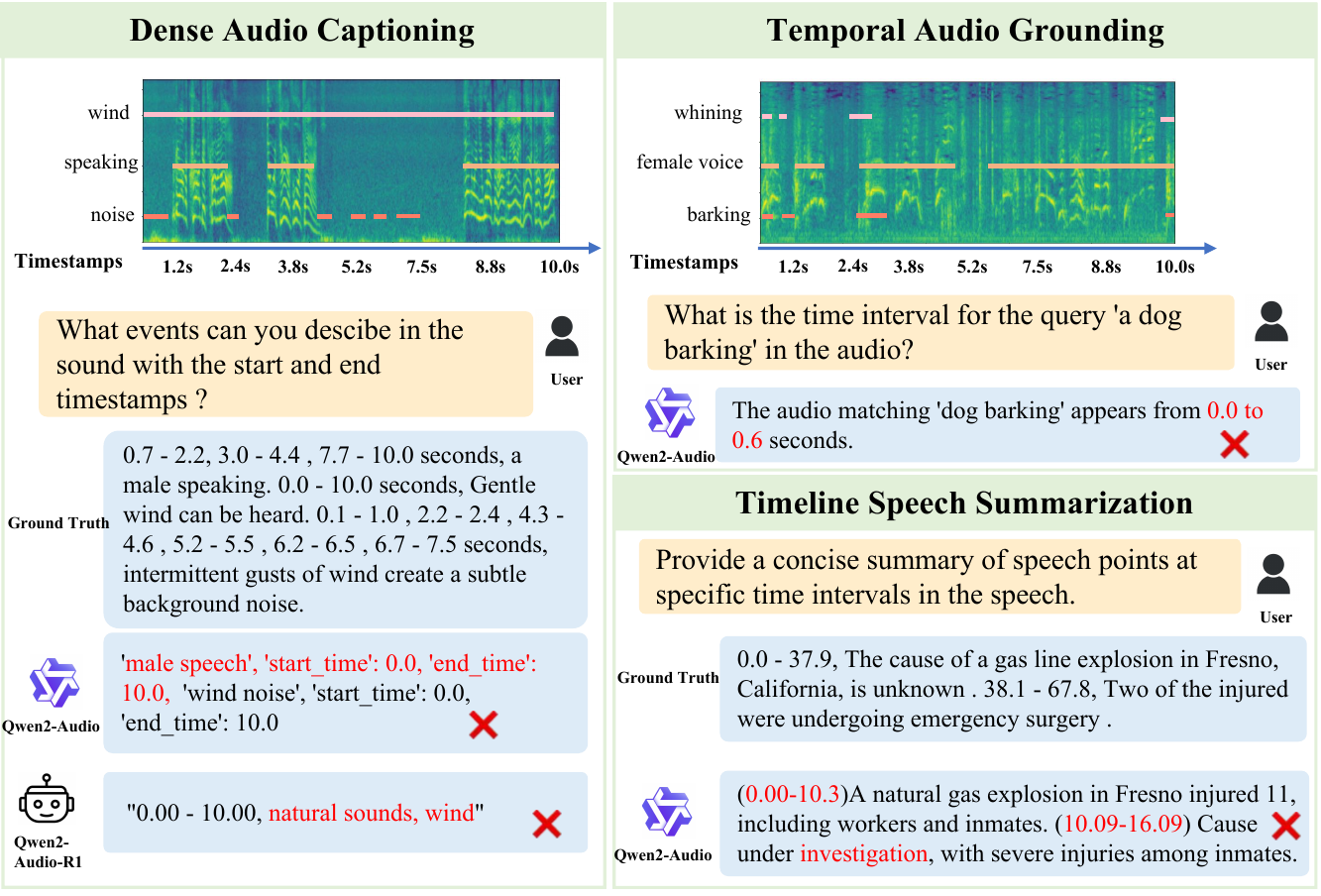}
    \caption{Example of failed cases by Qwen2-Audio and Qwen2-Audio-R1 on fine-grained tasks that require both semantics and timestamps as output.}
    \label{fig:bad}
\end{figure}
Audio, mainly including speech and non-speech sounds, is fundamental to human life, helping us perceive our surroundings, gather crucial information, and interact with others. 
To automatically interpret acoustic content and map it to human cognition, models are specially designed with advanced neural architectures and learning schemes to align audio features with natural language. 
Due to the intrinsic linguistic nature, speech tasks, such as automatic speech recognition and spoken question answering, have been easily integrated with language models \cite{chuang2019speechbert, zhang2023googleusm}. 
For environmental sound, methods like Contrastive Language Audio Pre-training (CLAP) \cite{wu2023clap} has been proposed to embed audio and its corresponding caption into a shared latent space. 
However, these models are restricted to fixed task formulations or exclusive audio types (either speech or sound), limiting their ability to achieve human‑like audio understanding. 

Leveraging rich knowledge in large language models (LLMs) \cite{achiam2023gpt}, large audio language models (LALMs), which integrate audio encoders into pre-trained decoder-based LLMs, enable free-form audio question answering (AQA) \cite{lipping2022clothoaqa} and unified audio understanding. For example, Qwen2-Audio \cite{chu2024qwen2} conducts large-scale pre-training to align audio and text modalities, followed by tuning on instruction data to further enhance the command-following capability. It not only demonstrates competitive performance on specific tasks compared to expert models (e.g., whisper-large-v3 \cite{radford2023whisper}) but also offers more flexible and natural interactions.

Despite the notable achievements witnessed in LALMs, they still fall short in fine-grained audio understanding \cite{xu2024towards, mesaros2021sound}, especially when precise timestamp prediction is required. To demonstrate this, we evaluate Qwen2-Audio's capability to summarize sound events or speech content along with the corresponding onset/offset. As shown in Figure~\ref{fig:bad}, it struggles to accurately link temporal locations with acoustic semantics or speech meanings. Moreover, it exhibits significant hallucination when processing long audio, as it lacks the capability for end-to-end comprehension of long-form audio. Similar poor performance from other LALMs can also be observed in the following experiments reported later (see Table~\ref{tab:main}). The underlying reasons may be two folds: (1) existing LALMs directly project audio features into the shared latent space without explicitly modeling detailed grounding information, making it difficult for the LLM decoder to predict precise timestamps; (2) the instruction tuning data and evaluation benchmarks \cite{sakshi2024mmau} focus on general audio understanding rather than highlighting fine-grained temporal reasoning.

To address the above shortcomings, we propose TimeAudio, a comprehensive framework that incorporates fine-grained acoustic cues into LALMs with enhanced module designs and a specially curated dataset. 
\begin{itemize} 
\item At the module level, temporal markers are integrated into LALM's vocabulary to reduce the convergence burden of numerical regression, and absolute time-aware encoding is adopted to explicitly inject timestep information into audio embeddings. Additionally, to efficiently handle long audio inputs, we devise a novel token selection and merging strategy, that balances token length with information density. 
\item At the data level, we first design several novel timestamps-related understanding tasks, such as dense audio captioning and timeline speech summarization, that require both high-level semantic understanding and low-level temporal grounding. Based on these tasks, a new large-scale instruction dataset named FTAR is constructed. We also introduce well-established metrics to evaluate LALMs' fine-grained understanding performance.
\end{itemize}
By incorporating the above techniques and being further fine-tuned on the gathered FTAR dataset, our proposed TimeAudio demonstrates significant improvements in temporal localization and fine-grained understanding capability compared to prior LALMs.

\begin{figure*}[!t]
    \centering
    \includegraphics[width=0.8\textwidth]{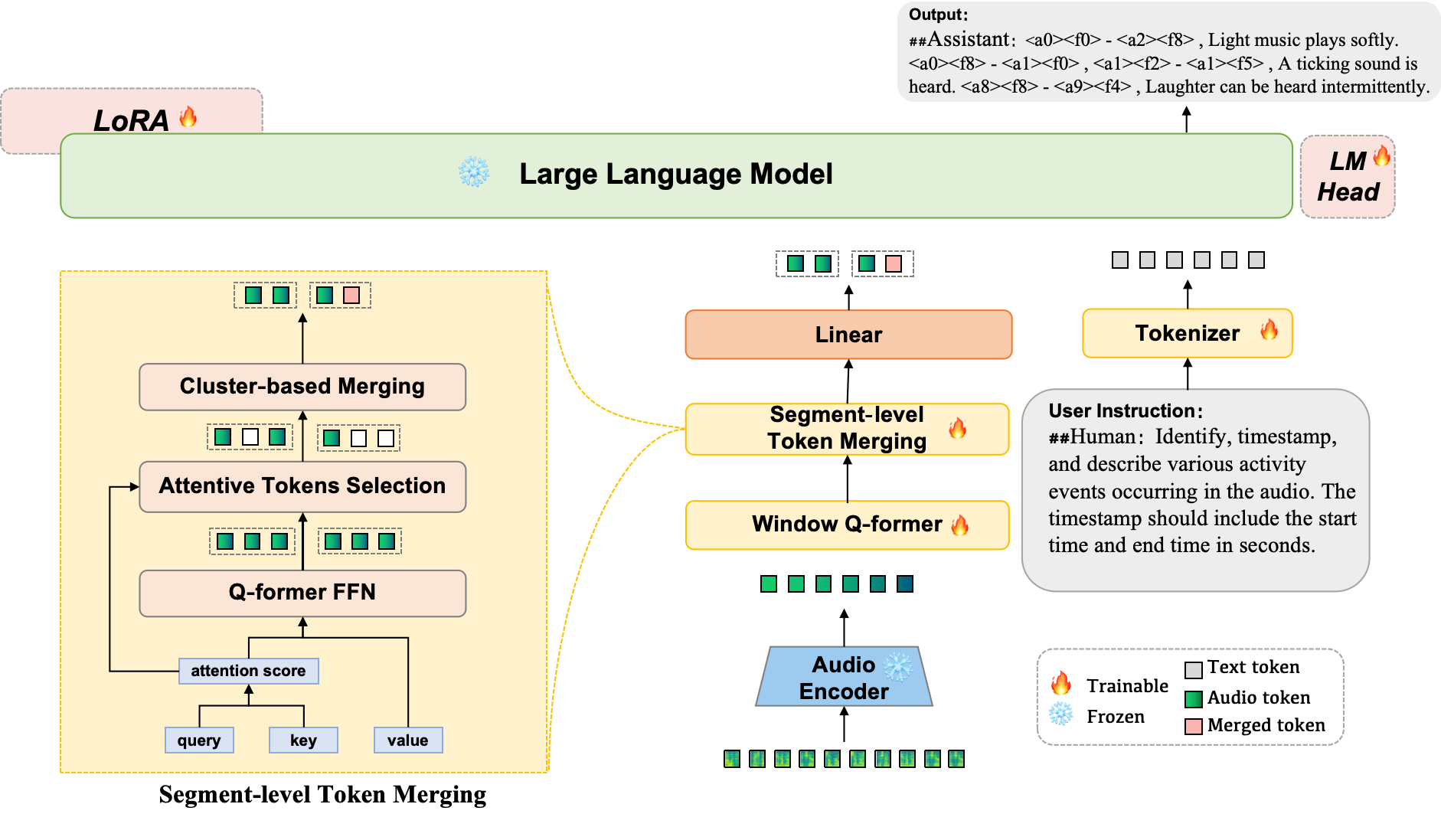}
    \caption{Overview of our TimeAudio method. The input audio is first split into segments and encoded into audio tokens. The window Q-former then projects these audio tokens into the language space and utilize a segment-level token merging to retain important semantic information along time. Timestamps are converted to special anchor and offset tokens. 
    \looseness=-1}
    \label{fig:overview}
\end{figure*}
\section{Related Work}
\subsection{Fine-grained Temporal Audio Understanding}
General audio understanding pays attention to overall content in the clip, for example, audio captioning \cite{wu2019audio} solely requires the model to output the contained sound events with simple temporal order descriptions. In contrast, fine-grained temporal understanding aims to grasp semantics along with their corresponding time intervals. For instance, audio grounding models \cite{xu2024towards} produce specific sound events with their timestamps, and some meeting summary tasks summarize key points with respective time spans\cite{hu2023meetingbank}. Fine-grained temporal understanding offers more traceable evidence and details for users, which is important to reduce hallucination and build responsible models. However, current LALMs perform unpromisingly regarding fine-grained temporal reasoning, while expert models \cite{wu2025flam, li2024mgaclap, wang2023leveraging} lack zero-shot capabilities, which limits their broader applications. Moreover, the absence of suitable instruction tuning datasets for temporal tasks, coupled with extreme task imbalance in existing datasets, hinders model performance. To address this issue, we construct a new, diverse dataset and propose TimeAudio, a model featuring time-sensitive modules that enhance LALM performance on fine-grained tasks.

\subsection{Large Audio Language Models}
Vanilla LLMs have proven effective at leveraging their captured knowledge for zero-shot solutions to general challenges. Recent research has explored to extend their capabilities further by integrating information from other modalities like video~\cite{liu2023visual} and audio~\cite{gong2023listen}. In the audio modality, LALMs project acoustic features into the embedding space of LLMs and leverage supervised finetuning to enhance the instruction following capability for audio inputs. The pioneer Pengi \cite{deshmukh2023pengi} utilizes fixed templates as prompts to align audio and text modality, showing promising outcomes compared to the original CLAP. LTU \cite{gong2023listen} and GAMA \cite{ghosh2024gama} enable the model to answer free-form sound-related questions by incorporating diverse AQA pairs and robust feature encoders. SALMONN \cite{tang2023salmonn} and Qwen2-Audio \cite{chu2024qwen2} further extend the paradigm to more audio types and instruction forms, significantly improving LALM's scalability. However, the previously mentioned LALMs frequently cause hallucination issues if provided with longer input, as they struggle to perform end-to-end processing on full-length audio. To handle longer inputs, Audio Flamingo2 \cite{ghosh2025af2} proposes a sliding window mechanism with CLAP to achieve long environmental sound comprehension. Qwen2.5-Omni \cite{xu2025qwen2} adopts a streaming transcription strategy to recognize full-length speech content. These methods are not directly transferable, as they depend on extensive training or specialized adaptations of the model architecture.

\section{Method}
In this section, we present TimeAudio, an LALM that utilizes temporal markers and incorporating two key modules: absolute time-aware encoding and segment-level merging. 
These modules aim to enhance TimeAudio's temporal awareness and expand its capacity to understand and localize details in audio tasks.
To further bridge the gap in fine-grained temporal reasoning and robust instruction following, we introduce the FTAR (fine-grained temporal audio reasoning) dataset -- a comprehensive dataset built for instruction tuning on time-sensitive tasks.
We then fine-tune our model on this dataset to fully unleash its capabilities.

\subsection{Overview of TimeAudio}
TimeAudio is based on the fundamental architecture of SALMONN~\cite{tang2023salmonn}. Its overview is provided in Figure~\ref{fig:overview}. Specifically, TimeAudio is consists of four components: a sliding audio encoder, a window Q-former, a segment-level token merging module, and an LLM to process raw audio. 
The sliding audio encoder first divides long audio into shorter segments and combines the BEATs~\cite{chen2022beats} and the Whisper encoder~\cite{radford2023robust} to extract features for each segments independently. Then, the window Q-former projects these encoded audio tokens into the language space and applies a segment-level token merging mechanism based on attention scores to filter out unimportant acoustic information.
Finally, the audio embeddings and the textual token embeddings of user prompts are fed into the LLM to generate response.

\subsection{Temporal Markers}
To improve LALM's comprehension and reduce its hallucination, it is promising to capture fine-grained temporal relationships instead of merely detecting the content of the audio.
Previous studies~\cite{sridhar2024enhancing, huang2024vtimellm} show that LLMs can inherently encode temporal information from sequential inputs, however, directly predicting precise timestamps over long audio remains challenging.  
In temporal audio understanding tasks, fine-grained audio comprehension requires temporal semantics across large time spans and during overlapping events.
However, predicting timestamps with the simple number token hinder the language model to capture precise semantics.
Furthermore, the relative time tokens~\cite{wang2024grounded, bain2023whisperx} with fixed intervals (\verb|<0.2>| \verb|<0.4>| \verb|<0.6>|...) imposes a heavy burden on the LLM's vocabulary and brings quantization errors in audio processing.

To resolve these issues, we introduce \textbf{\textit{Temporal Markers}}, which incorporate unique temporal tokens into the tokenizer to assist the LLM perceive specific timestamps. 
Given a fine-grained caption depicting a particular audio clip and its associated timestamps, we have designed anchor and offset tokens to convert continuous timestamps into a sequence of discrete temporal tokens.
The anchor token grounds the prediction immediately while the offset tokens represent a fine-grained adjustment. 
This strategy reduces the total number of temporal tokens, maintaining constant precision regardless of audio length.
We then convert the timestamp-related text into the unified temporal marker format.  
Textual and temporal tokens are mapped into a shared semantic space through the extended word embedding layer. An example input (containing male voice from 0.0s to 2.5s, 3.2s to 8.0s and soundtrack from 0.0s to 9.0s) is shown below:
\begin{tcolorbox}[
    colback = gray!6,
    colframe = black!60,      
    boxrule = 0.6pt,         
    rounded corners,          
    left = 4pt, right = 4pt,  
    top = 2pt, bottom = 2pt,
    fontupper = {\ttfamily\footnotesize} 
]
<s><audio>$\mathbf{F}_{audio}$</audio>\
<a0><f0> - <a2><f5>, <a3><f2> - <a8><f0>, A male voice delivers a great performance.
<a0><f0> - <a9><f0>, the soundtrack is filled with rich music.</s>
\end{tcolorbox}
\noindent where \verb|<s>| and \verb|</s>| indicate the start and end of sequence, \verb|<audio>| and \verb|</audio>| indicate the start and end of encoded audio features. \verb|<a>| and \verb|<f>| indicate the anchor and offset token with different time.
Despite solving quantization errors with temporal marker tokens, randomly initializing these tokens degrade the pretrained embedding space.
To tackle this issue, we transfer knowledge that implicitly contained in raw numeral to these anchor tokens, as temporal understanding is already implicitly contained within the LLM.
Furthermore, we compute the embedding for each offset token by averaging the embeddings of its numeral tokens and the decimal‐point token. For example, consider the time token ${\langle a_0\rangle}$, ${\langle f_0\rangle}$ and its embedding:
\begin{align}
[\mathbf{W}_{\text{token}}]_{\mathrm{ID}(\langle a_0\rangle)} &= [\mathbf{W}_{\text{token}}]_{\mathrm{ID}(0)} \notag\\
[\mathbf{W}_{\text{token}}]_{\mathrm{ID}(\langle f_0\rangle)} &= ({[\mathbf{W}_{\text{token}}]_{\mathrm{ID}(0)} + [\mathbf{W}_{\text{token}}]_{\mathrm{ID}(\text{.})}})/{2}
\end{align}
where ID() denotes the token ID of the input token. To ensure knowledge is properly transferred from numeral tokens to temporal tokens at the final prediction stage, we apply the same initialization to the LLM's prediction head.

\subsection{Absolute Time-aware Encoding}
\label{sec:Time-aware}
While recent multi-modal models~\cite{xu2025qwen2, guo2025vtg} have demonstrated the efficiency of absolute temporal positions, LALMs still suffer degraded performance when understanding the exact temporal order of events.
The considerable diversity in event and speech prosody makes it difficult for the model to accurately identify the true temporal locations and perform effective searching.
To enhance temporal awareness of the audio feature, we introduce a time-aware encoding that explicitly grounds acoustic features to absolute timeline.

Given a long audio $X$, we divide it into $N_s$ length segments, represented as $\mathbf{X}_a = \{x_i\}_{i=0}^{N_{s}-1}$. And each segment $x_i$ is encoded to a contiguous audio embedding as:
\begin{align}
\mathbf{W}_i = \text{Concat}(\mathcal{G}_\text{Whisper}({x}_i),\mathcal{G}_\text{BEATs}({x}_i))
\end{align}

\noindent where the dual audio encoders $\mathcal{G}(\cdot)$ extract speech and audio feature independently, and these are concatenated frame-by-frame along the feature dimension, yielding the acoustic embeddings $\mathbf{W}_i$. Then, we construct learnable the absolute time embedding $\mathbf{W}_{t}$ to explicitly provide information for temporal grounding. This preserves the sequence embedding with the relative order while accurately reflecting the specific position of each time point in the audio sequence. The audio embedding is augmented with its corresponding absolute time embedding:
\begin{align}
\mathbf{e}_t(t_i) &= \mathbf{h}(j_i)^{\boldsymbol{\top}}\mathbf{W}_t \;=\; [\mathbf{W}_t]_{j_i}\in\mathbb{R}^{d},\\
\hat{\mathbf{W}}_i &= \mathbf{W}_i + \mathbf{e}_t(t_i).
\end{align}
where $t$ represents the absolute timestamp (in seconds) associated with the corresponding segments. 
The time embedding is selected by one-hot lookup $\mathbf{h}(j_i)^{\top}\mathbf{W}_t=[\mathbf{W}_t]{j_i}$, and $j_i$ denotes the discretized time index.
The time embedding $\mathbf{W}_{t}$ is zero-initialized to preserve the integrity of the pretrained audio encoders during the initial phases of training. We posit that absolute time-aware and positional encodings are orthogonal, a claim validated by our experimental evidence.


\subsection{Segment-level Token Merging}
After obtaining the audio tokens with temporal feature, we apply the Q-former to project $T$-frames audio into $L$
semantic tokens. 
Although Q-Former uses a fixed number of queries for alignment, handling long audio is still costly in computation.
Some existing methods typically compress speech feature through higher-level projections~\cite{kang2024prompting} or temporal down-sampling~\cite{shang2024end}.
Inspired by VisionZip~\cite{yang2025visionzip}, we incorporate a segment-level token merging to prune the redundant tokens in the end-to-end structure. 
Our method reuses Q-former's attention information, avoiding extra computation and memory overhead. We discuss the details of the segment token merging as follows.
\paragraph{Attentive Tokens Selection.}
In speech summarization, the particular concern is that long speech usually contains extensive transcripts with dispersed information.
Therefore, our goal is to adaptively select important audio tokens from each audio segment and merge the redundant tokens into contextual feature.
In order to evaluate the relative importance of each audio token, we investigate the attention scores computed within the Q-Former. Specifically, we compute the attention matrix:
\begin{align}
\mathbf{A} &= \operatorname{Softmax} \!\bigl(\mathbf{Q}\mathbf{K}^{\mathsf T}/\sqrt{D}\bigr) \;\in\; \mathbb{R}^{B \times N_q \times N_q}
\end{align}
where $\mathbf{A}$ represents the multi-head attention scores, $D$ is the state dimension, and $\mathbf{Q}$ and $\mathbf{K}$ represent query and key from ${\mathbf{W}}_i$, respectively.
We identify the most salient tokens by averaging the attention scores across all heads to produce a unified score matrix. Tokens that receive higher average attention from all other tokens in the sequence are more significant and are therefore preserved.

\paragraph{Cluster-based Tokens Merging.}
After selecting the attentive audio tokens, we introduce a cluster-based merging process to retain information from the remaining tokens. 
This approach is based on the insight that the attention key vectors already capture the salient content of each token. 
To guide the merging, we first uniformly split the remaining tokens into target tokens and merge candidates.
Then, we compute a similarity metric between the key vectors of the target and candidate tokens:
\begin{align}
\mathbf{Sim}(\mathbf{h}_{i},\mathbf{h}_{j})= \mathbf{k}_{i}\,\mathbf{k}_{j}^{\mathsf T},
\end{align}
where $\mathbf{Sim}(\mathbf{h}_{i},\mathbf{h}_{j})$ is the similarity obtained by the dot product of the key vectors; $\mathbf{h}_{i,j\in\{1,\dots,n\}}$ denotes the remaining audio tokens.
The candidate token assigned to the centroid to which it exhibits the highest semantic similarity.
Finally, the most semantically-related tokens are fused together, yielding a set of contextual tokens.

\subsection{Training Strategy}
To develop an efficient LALM, we utilize a two-stage training pipeline as follows:
\paragraph{Stage-1: Temporal Token Alignment.}
In the first stage, we continue pre-training the model using checkpoints from well-trained LALMs to align fine-grained audio features with temporal information. 
We collect a wide range of grounding datasets, focusing on tasks such as Temporal Audio Grounding, Dense Audio Captioning, and Timeline Speech Summarization, that enable the model to search for and localize temporal information effectively.
To facilitate this, we make the LoRA adapters, window Q-former, and the absolute time and special text embeddings trainable, while keeping the audio encoder and LLM frozen.

\paragraph{Stage-2: Long Audio Instruction Tuning.}
While the initial pre-training stage equips the model with a foundational capability for temporal reasoning, a significant mismatch occurs when it confronts long audio sequences, often resulting in semantic misalignment. 
To bridge this gap, the second stage focuses on enhancing its ability to understand long speech and respond to diverse instructions. 
This is achieved by fine-tuning the window Q-former and LoRA on a small set of instruction data, while keeping all other components frozen. 
This stage improves long audio comprehension while aligning both the temporal knowledge and the LLM's semantic space through diverse instructions.

\begin{figure}
    \centering
    \includegraphics[width=1.\linewidth]{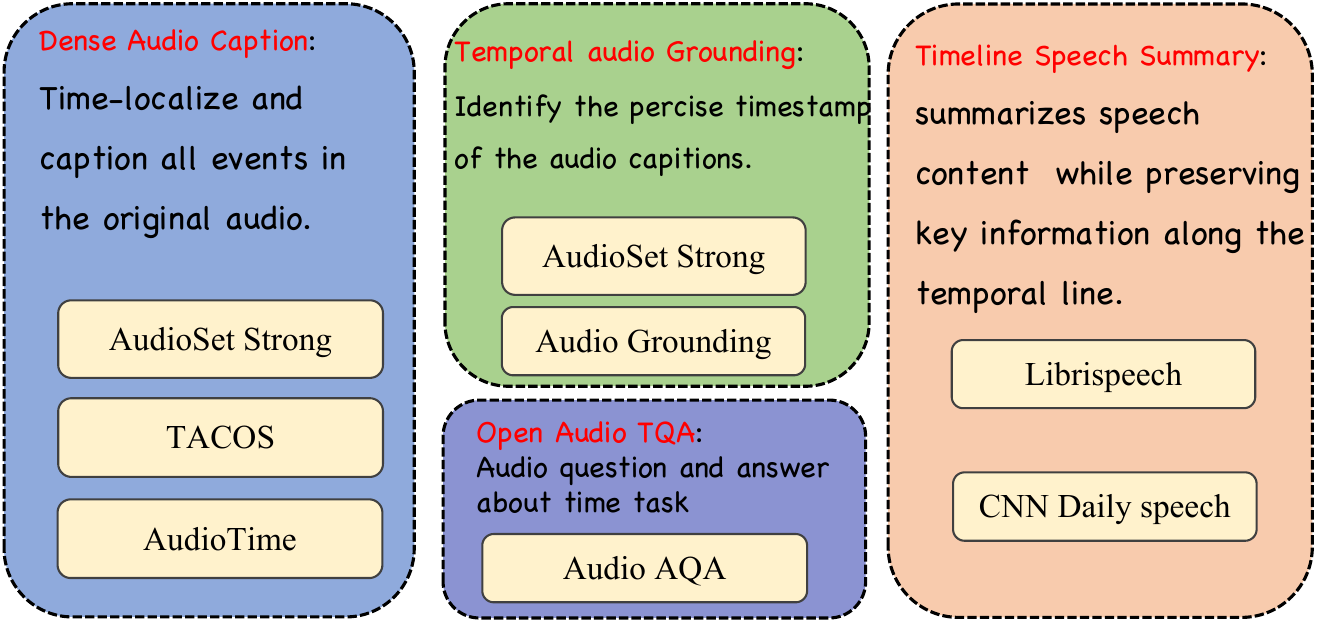}
    \caption{Involved tasks and datasets in the time-aware instruction tuning dataset.}
    \label{fig:dataset}
\end{figure}

\subsection{FTAR Dataset Construction}
In this section, we introduce the FTAR, a dataset comprising 260K publicly available audio-text pairs.
It is composed of three audio tasks centered on temporal reasoning, drawn from diverse domains as detailed in Figure~\ref{fig:dataset}.
The FTAR Dataset is specifically designed to help users obtain fine-grained content during interactions with AI assistants. Additional details are available in Appendix A.

\begin{itemize}
    \item \textbf{Dense Audio Captioning:} The task of dense audio captioning requires generating event descriptions with their respective start and end times, formatted as: \verb|<start>|-\verb|<end>|, \verb|captions|. We employ the Qwen2.5 model~\cite{team2024qwen2} to refines coarse-grained event labels into rich, fine-grained descriptions. We utilize real-world AudioSet-Strong~\cite{hershey2021benefit}, TACOS~\cite{primus2025tacos}, AudioTime~\cite{xie2025audiotime} dataset to construct the task. 

    \item \textbf{Temporal Audio Grounding:} The temporal audio grounding task is defined as localizing a specific event within an audio based on a descriptive sentence. For this task, we treat the event caption as a query and output the corresponding start and end times in the format \verb|<start>|-\verb|<end>|. We collect AudioSet-Strong, AudioGrounding~\cite{xu2024towards} and formulate them.

    \item \textbf{Timeline Speech Summarization:} For the speech summarization task, it aims to condense spoken content while preserving key information along the temporal line, emphasizing both high-level semantic extraction and low-level grounding ability. We utilize F5-TTS ~\cite{chen2024f5} to synthesize speech from CNN/DailyMail summarization~\cite{nallapati2016abstractive}. Beyond synthetic data, We further collect Librispeech~\cite{panayotov2015librispeech} segments for the task.
    
\end{itemize}
To enhance a rich diversity of temporal instructions, the Audio Temporal Question Answer (TQA) task is constructed from the OpenAQA dataset~\cite{gong2023listen}, which consists of free-form and diverse audio question-answer pairs on counting, duration, and time sequence. We also include audio captioning and speech recognition data to retain model's general audio understanding capabilities.

Table~\ref{tab:datasets} presents the instruction tuning data across different tasks, highlighting the broad coverage in data scale, task coverage, and audio durations.

\section{Experiments}
\begin{table}[!t]
    \centering
    \small
    \setlength{\tabcolsep}{1.pt}
    \begin{tabular}{lccc}
    \toprule
    Tasks & \# Sub-data & \# Samples & Avg. len \\
    \midrule
    Dense Audio Captioning & 3 & 110K & 11.3s \\
    Temporal Audio Grounding & 2 & 100K & 9.8s \\
    Timeline Speech Summarization & 2 & 42K & 81.7s \\
    Audio TQA & 1 & 15K & 10.0s \\
    \bottomrule
\end{tabular}
\caption{Datasets Used in Training for Various Tasks.}
\label{tab:datasets}
\end{table}

\subsection{Evaluation Setups}

\begin{table*}[t]
\renewcommand\arraystretch{0.95}
\setlength{\tabcolsep}{0.5pt}
\small
\centering
\begin{tabular}{cccccccccccc}
\toprule
\multirow{2}{*}{\textbf{Model}} & \multirow{2}{*}{\textbf{LLM Scale}} & \multicolumn{3}{c}{\textbf{Dense Audio Captioning}} & \multicolumn{4}{c}{\textbf{Temporal Audio Grounding}} & \multicolumn{3}{c}{\textbf{Timeline Speech Summarization}} \\
\cmidrule(lr){3-5} \cmidrule(lr){6-9} \cmidrule(lr){10-12}
& & METEOR & Eb-F1 & At-F1 & R@0.5 & R@0.7 & R@0.9 & mIoU & ROUGE-1 & ROUGE-L & mIoU \\
\midrule
\multicolumn{12}{c}{\textit{Zero-shot LALMs}} \\
\midrule
Qwen2-Audio~\cite{chu2024qwen2} & 7B & 6.7 & 9.8 & 50.3 & 32.1 & 18.7 & 10.8 & 20.5 & 17.4 & 12.3 & 13.3 \\
Qwen2.5 omni~\cite{xu2025qwen2} & 7B & 3.4 & 9.6 & 36.0 & 17.9 & 10.9 & 6.5 & 11.8 & 15.6 & 11.7 & 12.8 \\
Qwen2-Audio-R1~\cite{li2025reinforcement} & 7B & 3.7 & 8.5 & 40.6 & 29.3 & 17.3 & 8.8 & 18.5 & 15.2 & 11.5 & 16.2 \\
Hubert-MiniChat~\cite{kang2024prompting} & 3B & - & - & - & - & - & - & - & 33.6 & 22.3 & - \\
\midrule
\multicolumn{12}{c}{\textit{FTAR-Tuned LALMs}} \\
\midrule
GAMA~\cite{ghosh2024gama} & 7B & 19.7 & 11.0 & 67.8 & 31.0 & 21.1 & 14.8 & 22.3 & - & - & - \\
Qwen audio~\cite{chu2023qwen} & 7B & 10.6 & 13.5 & 43.6 & 53.4 & 32.2 & 23.7 & 36.4 & 15.9 & 20.5 & 76.8 \\
Qwen2-Audio~\cite{chu2024qwen2} & 7B & \textbf{22.4} & 36.5 & 67.8 & 72.8 & 55.4 & 26.8 & 51.7 & 40.0 & 28.5 & 85.2 \\
SALMONN-7B~\cite{tang2023salmonn} & 7B & 19.9 & 32.4 & 68.0 & 71.4 & 55.8 & 28.6 & 51.9 & 39.5 & 28.7 & 84.3 \\
SALMONN-13B~\cite{tang2023salmonn} & 13B & 20.2 & 32.0 & 67.6 & 69.2 & 53.5 & 28.3 & 50.3 & 40.2 & 29.8 & 88.2 \\
\midrule
TimeAudio (ours) & 7B & 20.4 & \textbf{37.4} & \textbf{70.5} & \textbf{75.7} & \textbf{61.2} & \textbf{36.5} & \textbf{57.8} & \textbf{42.4} & \textbf{30.8} & \textbf{94.2} \\
\bottomrule
\end{tabular}
\caption{{Comparison of performance on the fine-grained temporal task with other LALMs' methods. The \textbf{bold} item denotes the best result.}}
\label{tab:main}
\end{table*}

\paragraph{Implementation Details.} 
We use 7B SALMONN as the base LALM for experiments~\cite{tang2023salmonn}. The window Q-former module and the weights of sequence embedding are initialized using the pre-trained checkpoint. We augment its vocabulary with M=20 specialized temporal tokens. The learnable time embedding is configured with a maximum of 768 positions.
During data pre-processing, each audio is split into a sequence of 30-second intervals and the max number of segments is 5. 
Furthermore, to improve computational efficiency, we retain only 22 attentive tokens and 4 contextual tokens, effectively pruning 75\% of redundant information. Our two-stage training process involves 10 epochs of continual pre-training using a learning rate of 1e-5 , followed by 5 epochs of instruction tuning with a learning rate of 2e-6. More details can be found in Appendix B.

\paragraph{Evaluation Tasks and Metrics.} 
For dense audio captioning, we test on the AudioSet-Strong evaluation set~\cite{hershey2021benefit} with caption less than 100 words. Metrics such as METEOR score~\cite{banerjee2005meteor}, event-based measures (Eb) and clip-level macro F1 score (At)~\cite{mesaros2016metrics} are applied to evaluate the time perceptive ability and the diversity of descriptions between the generated events and the ground-truth.
For temporal audio grounding, we evaluate on the AudioGrounding test data and report the mean Intersection over Union (mIoU) and Recall at IoU thresholds of {0.5, 0.7, 0.9} between predicted and ground-truth timestamps.
To evaluate timeline speech summarization, we follow the protocol of~\cite{kang2024prompting} and use a subset of the CNN/DailyMail test set, including only articles under 1600 characters. We report standard content quality metrics (ROUGE~\cite{lin2004rouge}) and supplement them with a mean Intersection over Union (mIoU) score to measure how accurately the summary is grounded to the audio timeline. More details are available in Appendix C.

\paragraph{Compared Methods.} 
For our baseline evaluation, we first represent general-purpose LALMs used in a zero-shot inference performance. We select Qwen2-Audio~\cite{chu2024qwen2}, Qwen2-Audio-R1~\cite{li2025reinforcement}, and Qwen2.5 Omni~\cite{xu2025qwen2}. 
Second, for models specifically adapted for fine-grained tasks, we evaluate several prominent audio LLMs, including Qwen Audio~\cite{chu2023qwen}, Qwen2-Audio, GAMA~\cite{ghosh2024gama}, and SALMONN~\cite{tang2023salmonn}, all of which undergo parameter-efficient fine-tuning.

\subsection{Main Results}
\paragraph{Dense Audio Captioning.} This task involves accurately capturing the temporal locations of all sound events within an audio clip, alongside providing faithful descriptions that match the underlying time.
As shown in Table~\ref{tab:main}, existing LALMs exhibit significant limitations in precise temporal localization under zero-shot conditions, a fact underscored by the top-performing Qwen2-Audio, which achieves an Eb-F1 score of 9.8.
The inaccurate event localization directly impacts the captioning evaluation, such as the METEOR metrics. 
Our method exhibits a clear strength in temporal accuracy, achieving a remarkable performance gain over fine-tuned Qwen2-Audio with +0.9 Eb-F1 and +2.7 At-F1 scores. 
This demonstrates that TimeAudio effectively process audio with precise event localization.
As for the METEOR score, the fine-tuned Qwen2-Audio model achieves even stronger performance. 
We speculate that this is due to the pre-trained checkpoint used for initialization, which not being trained on sufficiently diverse audio captioning.

\paragraph{Temporal Audio Grounding.} This task directly reflects the ability to precisely localize time interval with a given query event. Results show that TimeAudio achieves 57.8 score on mIoU of the AudioGrounding dataset, which surpasses the fine-tuned LALMs, i.e. SALMONN-7B, by a substantial 11.4\% gain in mIoU.
This demonstrates a superior overall accuracy in fine-grained temporal perception for given textual descriptions.
It is particularly noteworthy that TimeAudio achieves its greatest performance gains on the temporal audio grounding task. We argue that the substantial improvement, centered on a task that explicitly measures temporal localization, demonstrates the effectiveness of our proposed methods.

\paragraph{Timeline Speech Summarization.} 
While the dense audio captioning and temporal audio grounding focus on audio tasks, this task aims to enhance fine-grained speech understanding at the segment level.
Overall, our model achieves a 42.4 ROUGE-1 and 30.8 ROUGE-L on CNN/DailyMail speech, outperforming other LALMs with speech summarization capabilities.
This highlights the contribution of our segment-level token merging in retain the important semantics of each audio segment. 
Besides, TimeAudio surpasses even specialized, task-specific Hubert-MiniChat, which demonstrates both the challenging nature of the task and the superiority of our model in processing long audio.

\begin{table*}[t]
\renewcommand\arraystretch{0.95}
    \begin{center}
    \small
    \begin{tabular}{l c c c c c c c c c c}
        \toprule
        \multirow{3}{*}{\textbf{Method}} & \multicolumn{3}{c}{\textbf{Dense Audio Captioning}} & \multicolumn{4}{c}{\textbf{Temporal Audio Grounding}} & \multicolumn{3}{c}{\textbf{Timeline Speech Summarization}} \\
        \cmidrule(lr){2-4} \cmidrule(lr){5-8} \cmidrule(lr){9-11} 
        & METEOR & Eb-F1 & At-F1 & R@0.5 & R@0.7 & R@0.9 & mIoU & ROUGE-1 & ROUGE-L & mIoU\\ 
     
        \midrule
        SALMONN (FTAR-Tuned) & 19.9 & 32.4 & 68.0 & 71.4 & 55.8 & 28.6 & 51.9 & 39.5 & 28.7 & 84.3 \\
        \midrule
        \quad +w/ TM (not initialize) & 19.5 & 33.5 & 69.2 & 71.7 & 56.2 & 33.5 & 52.3 & 40.0 & 28.8 & 87.5 \\
        \quad +w/ TM & 20.5 & 36.0 & 70.8 & 72.7 & 57.8 & 34.2 & 54.9 & 41.0 & 29.8 & 88.7 \\
        \quad +w/ ATE & 20.3 & 35.8 & 69.5  & 72.0 & 56.9 & 32.5  & 53.8 & 40.6 & 30.1 & 86.6 \\
        \quad +w/ TM + ATE & \textbf{20.8} & \textbf{37.8} & \textbf{71.4} & 73.7 & 58.6 & 35.5 & 56.0 & 41.4 & 29.8 & 90.4 \\
        \quad +w/ TM + ATE + SEM & 20.4 & {37.4} & {70.5} & \textbf{75.7} & \textbf{61.2} & \textbf{36.5} & \textbf{57.8} & \textbf{42.4} & \textbf{30.8} & \textbf{94.2}\\
        \bottomrule
    \end{tabular}
    \caption{\textbf{Ablation study on different module components}. We analyze and compare the effects of using time-sensitive modules for TimeAudio. TM indicates Temporal Marker, ATE indicates Absolute Time-aware Encoding and SEM indicates Segment-level Token Merging. The \textbf{bold} item denotes the best result.}
    \label{tab:ablation_performance}
\end{center}
\end{table*}

\begin{table}[t]
  \centering
  \small
  \vspace{-2em}
  \begin{tabular}{@{}cccc@{}}
    \toprule
    \multirow{2}{*}{\textbf{Retained Ratio}} & \multicolumn{3}{c}{\textbf{Timeline Speech Summarization}} \\
    \cmidrule(lr){2-4}
    & {ROUGE-1} & {ROUGE-L} & {mIoU} \\
    \midrule
    0.10 & 24.3 & 18.4 & 72.9  \\
    0.15 & 31.6 & 20.5 & 73.2 \\
    0.20 & 40.2 & 29.7 & 84.8 \\
    \rowcolor{gray!12}
    \textbf{0.25} & {42.4} & {30.8} & \textbf{94.2}   \\
    0.30  & {42.5} & 31.1 & {94.0}  \\
    \bottomrule
  \end{tabular}
   \caption{Ablation for dominant audio tokens retained ratio.}
    \label{tab:ablation}
\end{table}

\subsection{Ablation Study}
In this section, we systematically conduct ablation studies on TimeAudio.
Specifically, we evaluate the effectiveness of our proposed module (excluding data effects) in Table~\ref{tab:ablation_performance} and analyze the impact of the token merging in Table~\ref{tab:ablation}.
\paragraph{Multi-Task Ablation.}
The TM-only setup yields substantial improvements on time-sensitive tasks, with gains of +3.6 Eb-F1 on AudioSet-Strong, +3.0 mIoU on AudioGrounding and +4.4 mIoU on CNN/DailyMail.
These findings suggest that using time tokens significantly enhanced temporal understanding to accurately locate the timestamps.
In the case of adding TM and ATE, the model’s ability to temporally describe audio is enhanced, as indicated by an increase of 0.9 in METEOR and 5.4 in the Eb-F1.
Adding segment-level token merging introduces expected trade-offs: while it improves timeline speech summarization accuracy on ROUGE-1 and mIOU, it leads to a moderate decline in dense captioning precision. 
The STM method saves more semantic information in long-form audio and improve alignment between the summarization and the audio content. 
These results highlight the effectiveness of our novel modules in the TimeAudio architecture.

\paragraph{Performance with More Attentive Tokens.}
As illustrated in Table ~\ref{tab:ablation}, we present the results of token merging method under different ratio of retained tokens. 
A smaller ratio of retained tokens fails to capture sufficient semantic information across long audio, resulting in suboptimal performance in both aspects. 
Conversely, raising the attentive-token count offers marginal improvements in semantic understanding before the gains level off.
We finally set the ratio of attentive tokens to 0.25 as a balanced trade-off between performance and computational efficiency.

\subsection{Qualitative Performance}
\begin{figure}[!t]
    \centering
    \vspace{-2em}
    \includegraphics[width=0.7\linewidth]{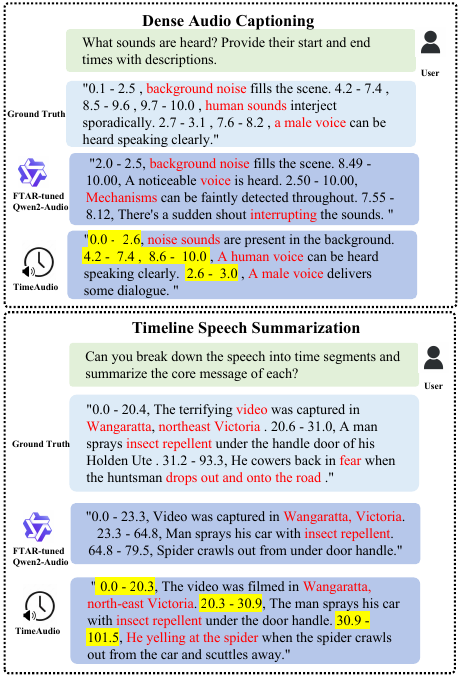}
    \caption{Qualitative results between models on the dense audio captioning task and timeline speech summarization task. Yellow denotes the time interval and red marks the audio content.}
    \label{fig:case}
\end{figure}

In Figure~\ref{fig:case}, we compare our method with ground-truth and Qwen2-Audio to highlight TimeAudio’s superior temporal awareness and understanding.
The first example presents a multi-event changes between short audio.
While Qwen2-Audio model exhibits a tendency to predict delayed start times and struggle to identify the complete time interval, our method demonstrates superior precision by accurately capturing the correct relative boundaries of the event.
The second case features a long audio characterized by events that happen in a specific order on CNN/DailyMail speech. 
A key limitation of the Qwen2-Audio model is its short context input, which necessitates the ability of LLM to guess contents in longer audio segments. 
These qualitative examples underscore the robustness and precision of our method in scenarios that are especially challenging for other base methods. We provide additional cases on these tasks in Appendix D.

\section{Conclusion}
In this work, we propose TimeAudio, a time-sensitive LALM capable of fine-grained perception and understanding.
This is enabled by a novel model design that incorporates temporal markers and absolute time-aware encoding for effective temporal modeling. 
To ensure effective compression in long audio, we introduce a segment-level token merging method to progressively preserves dominant information.
Additionally, we construct the FTAR dataset to further strengthen the model’s temporal reasoning. 
Extensive experiments demonstrate that TimeAudio significantly improves performance in time-centric scenarios, outperforming baselines on downstream temporal audio tasks while retaining general audio comprehension. 
These results underscore the importance of time as a crucial dimension for fine-grained audio perceptive capability.


\section{Acknowledgments}
This work is supported by the National Natural Science Foundation of China (62276250), the National Key R\&D Program of China (2022YFF1203303) and key R\&D program of Ningxia Autonomous Region (2024FRD05068).
And also supported by the Major Project of the National Social Science Foundation of China (21\&ZD292) and sponsored by Doubao Fund.
\bibliography{aaai2026}

\newpage
\appendix
\section{Appendix}
\subsection{Appendix A: FTAR Dataset Construction}

FTAR dataset contains 4 long timestamp-related audio tasks and incorporates 8 specific datasets derived from different domains. 
\paragraph{Dense Audio Captioning.} This task jointly addresses sound event localization and event captioning. The model is required to detect all constituent events within an audio clip and output a set of timestamped descriptions.
We gather AudioSet-Strong~\cite{hershey2021benefit}, TACOS~\cite{primus2025tacos}, AudioTime~\cite{xie2025audiotime} datasets to facilitate the understanding of significant events and sequence of occurrence. The AudioSet-Strong contains 97315 effective annotated audio with frame-level timestamp for 456 sound events. The TACOS include 12,358 audio recordings
from freesound with time annotations and short captions. AudioTime dataset is constructed by segments curation, sound
simulation, and captions generation with detailed temporal attributes as supervised signals, containing 5000 samples. 

\paragraph{Temporal Audio Grounding.} In this task, the model is given a natural language query as input and is required to output the corresponding temporal boundary, defined by a start and end time, within the audio. We include AudioSet-Strong, AudioGrounding~\cite{xu2024towards} datasets to achieve precise time localization when users query the event with natural captions. The AudioGrounding is a subset of AudioCaps augmented with human-annotated onsets and offsets of phrases, containing 4662 samples. 

\paragraph{Timeline Speech Summarization.} The goal is to condense spoken content while preserving essential information and its temporal context.
We utilize F5-TTS ~\cite{chen2024f5} to synthesize speech from CNN/DailyMail summarization~\cite{nallapati2016abstractive}, with 43018 samples. Librispeech~\cite{panayotov2015librispeech} segments are merged to achieve an long audio part.

\paragraph{Audio Temporal Question Answer (TQA).} This task provides a rich collection of free-form and open-ended audio question-answer pairs specifically designed to test reasoning about counting, duration, and time sequences, which is useful for instructional audios. We select 15k samples from OpenAQA dataset~\cite{gong2023listen} to obtain high-quality samples to enhance diversity.

\paragraph{Step I: Instruction Construction.} To ensure instruction quality and diversity, we follow a three-step process. We first manually write base instructions for each task. Then, we use Gemini 2.5 to generate more diverse expressions from these seeds. Finally, we manually select and refine the LLM outputs to create the final set of six high-quality instructions per task.

\paragraph{Step II: Answer Generation.} We leverage the Qwen2.5 model to refine coarse-grained event labels into rich, fine-grained descriptions, which are then timestamped in accordance with the user prompt. The overall quality of the FTAR data is ensured through the manual collection and curation of all constituent audio datasets.

\begin{tcolorbox}[
    colback = gray!6,
    colframe = black!60,      
    boxrule = 0.8pt,         
    rounded corners,          
    left = 4pt, right = 4pt,  
    top = 2pt, bottom = 2pt,
    fontupper = \ttfamily     
]
Transform the given sentence and keep the time ranges unchanged as they are in the format start - end seconds and multiple times. The caption should creatively reflect the events while maintaining accuracy and a natural flow. Multiple times should be placed before the respective description.
\end{tcolorbox}

\begin{table}[h]
\centering
\renewcommand{\arraystretch}{1.2} 
\caption{Hyper-parameter settings used in our experiments.}
\label{tab:hyperparameters}
\begin{tabular}{lc}
\toprule
\textbf{Hyper-parameter} & \textbf{Value} \\
\midrule
Sample rate & 16k \\
Max duration & 120s \\
\midrule
Stage 1 fine-tuning epochs & 10 \\
Stage 1 Batch size & 48 \\
Stage 1 Learning rate & 3e-5 \\
Stage 1 Warm-up learning rate & 1e-6 \\
Stage 2 fine-tuning epochs & 3 \\
Stage 2 Batch size & 24 \\
Stage 2 Learning rate & 1e-5 \\
Stage 2 Warm-up learning rate & 1e-7 \\
lora rank & 8 \\
lora alpha & 32 \\
lora dropout & 0.1 \\
Weight decay & 0.05 \\
AdamW $\beta$ & (0.9, 0.999) \\
\midrule
Second stride  & 0.33 \\
Second per window & 0.33 \\
Dominant token num & 24 \\
Contextual token num & 6 \\
Max text length & 300 \\
\midrule
Number of layers in win Q-Former & 6 \\
Hidden size of audio Q-Former ($D_Q$) & 768 \\
Hidden size of Vicuna-1.5 ($D_{LLM}$) & 4096 \\
\bottomrule
\end{tabular}
\end{table}

Table~\ref{tab:task_examples} provides a detailed information of the tasks included in our dataset, showcasing examples of the instructions, output formats, and expected responses.

\begin{table*}[t]
\centering
\renewcommand{\arraystretch}{1.4} 
\caption{Examples of instructions, output formats, and outputs for various fine-grained understanding tasks.}
\label{tab:task_examples}
\begin{tabular}{>{\raggedright\arraybackslash}p{3cm} p{12cm}}
\toprule
\textbf{Key} & \textbf{Value} \\
\midrule
\multicolumn{2}{l}{\textbf{\large Dense Audio Captioning}} \\
\midrule
Instruction Example & Which sound events occur, and what are their time intervals and descriptions? \\
Output Format & \verb|<timestamp_start>| - {\verb|<timestamp_end>|} seconds, {\verb|<description>|}, \dots \\
Output Example & 1.4 - 3.7, 4.3 - 5.1, 5.4 - 6.3, 7.0 - 8.7 seconds, A baby's cries pierce through the air intermittently. 2.5 - 2.9, 4.3 - 4.5 seconds, A young child speaks briefly. 0.3 - 10.0 seconds, Conversations fill the space. 0.2 - 0.4 seconds, A cough interrupts the ongoing dialogue.
 \\
\midrule

\multicolumn{2}{l}{\textbf{\large Temporal Audio Grounding}} \\
\midrule
Instruction Example & What are the start and end times of audio matching {\verb|<query_placeholder>|}? \\
Output Format & The given query happens in {\verb|<timestamp_start>|} - {\verb|<timestamp_end>|} seconds, \dots \\
Output Example & The given query happens in 0.2 - 2.7, 3.3 - 5.7, 6.5 - 7.2 seconds. \\
\midrule
\multicolumn{2}{l}{\textbf{\large Timeline Speech Summarization}} \\
\midrule
Instruction Example & Can you provide the speech summary with key points at specific time intervals? \\
Output Format & \verb|<timestamp_start>| - {\verb|<timestamp_end>|} seconds, {\verb|<summarization>|}, \dots \\
Output Example & 0.0 - 16.4, Vice Adm. Jan Tighe takes over as head of  U.S. Fleet Cyber Command, U.S. 10th Fleet. 16.4 - 84.5, She succeeds Adm. Michael Rogers, who moved on to become the NSA director. 84.5 - 120.5, The Navy, other branches have faced criticism for the treatment of female personnel.\\
\midrule

\multicolumn{2}{l}{\textbf{\large Audio TQA}} \\
\midrule
Instruction Example &  What is the duration of the background noise in the audio clip? \\
Output Example & The background noise in the audio clip lasts for the entire duration of 10.\\
Instruction Example &  What can be inferred about the number of men singing based on the duration and timing of the male singing sound events?\\
Output Example & Based on the duration and timing of the male singing sound events, it can be inferred that there are fibe men singing in this musical piece.\\
\bottomrule
\end{tabular}
\end{table*}

\subsection{Appendix B:Hyper-parameters for Instruction Tuning}
Tab.~\ref{tab:hyperparameters} lists full hyper-parameters for fine-tuning. The Hyper-parameters show our training details.

\begin{figure}[ht!]
\centering
\includegraphics[width=0.9\linewidth]{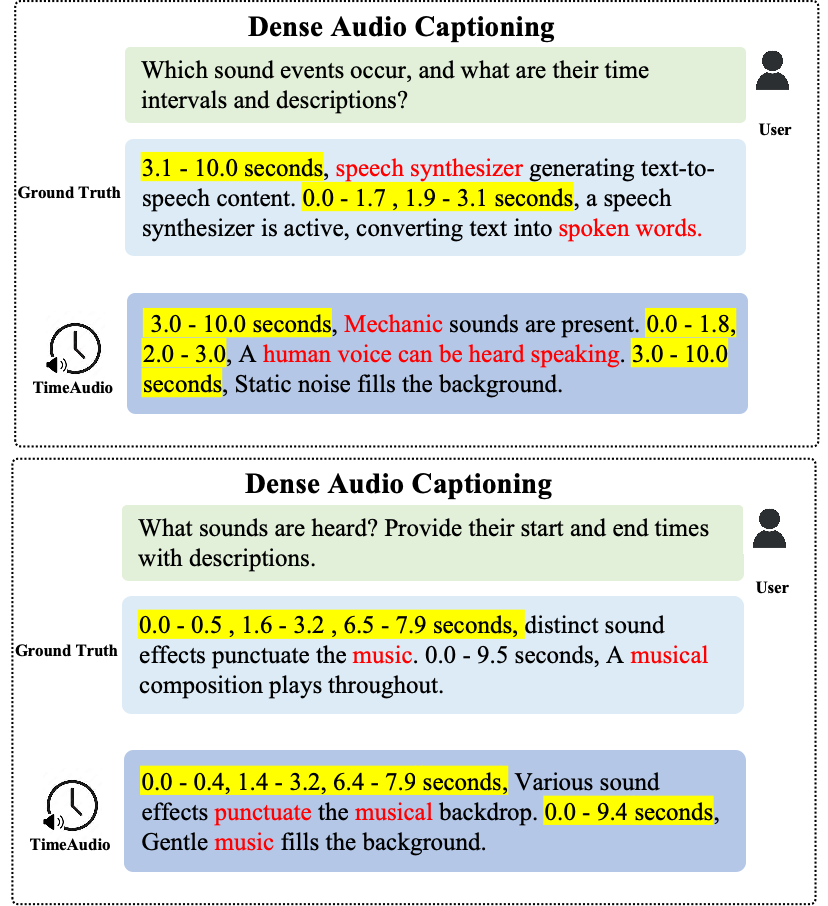}
\caption{Qualitative results on dense audio captioning task. For each audio, we ask audio to detect a series of events in the given audio and output the corresponding timestamps and descriptions.
}
\label{fig:dac}
\end{figure}

\begin{figure}[ht!]
\centering
\includegraphics[width=0.9\linewidth]{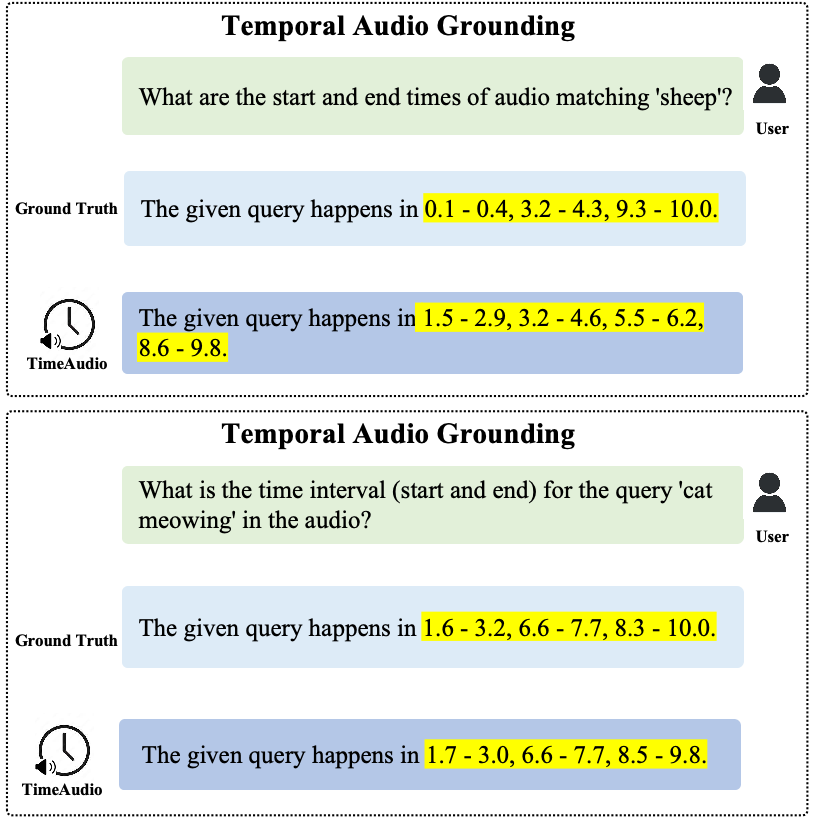}
\caption{Qualitative results on temporal audio grounding task.
}
\label{fig:tag}
\end{figure}

\begin{figure}[ht!]
\centering
\includegraphics[width=0.9\linewidth]{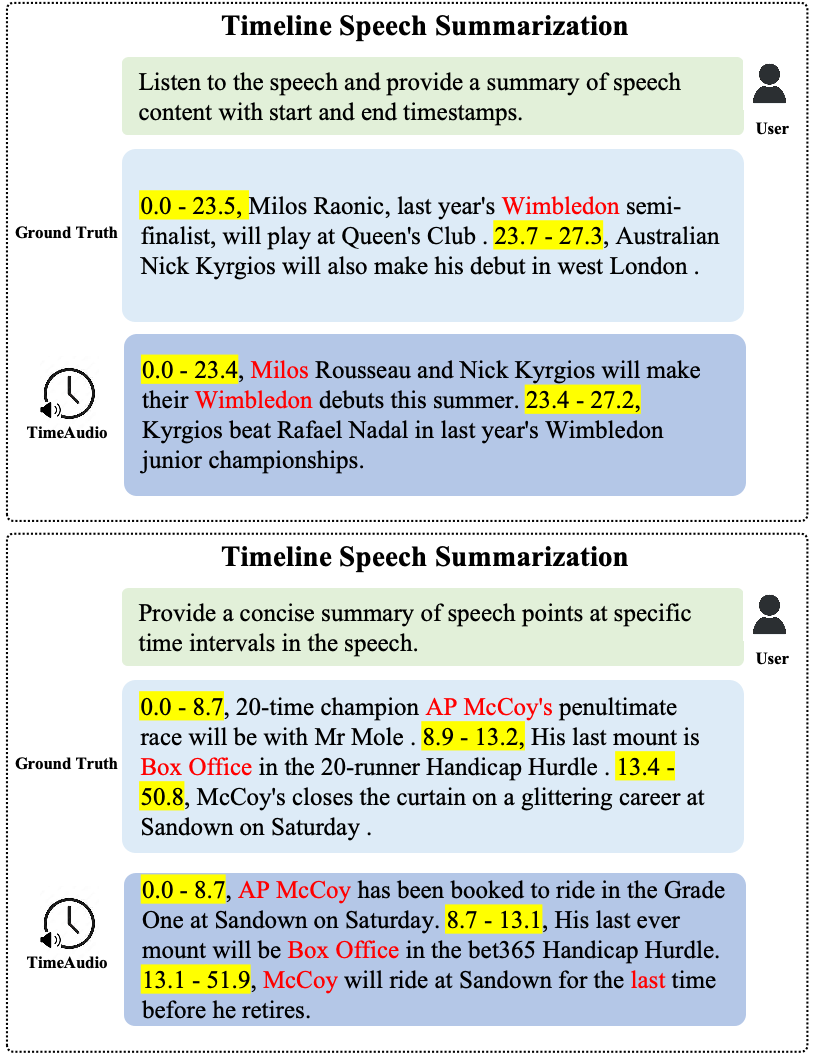}
\caption{Qualitative results on timeline speech summarization task. 
}
\label{fig:tss}
\end{figure}

\subsection{Appendix C: Details of Evaluation Datasets and Metrics}
\label{sec:evaluation}
To assess the model's zero-shot generalization, we evaluate its performance on a selection of practical and representative audio understanding benchmarks. These include dense audio captioning, temporal audio grounding, and timeline speech summarization.

\textbf{For dense audio captioning}, we use the part AudioSet-Strong evaluation dataset~\cite{hershey2021benefit}, which has 770 samples of sound events. 
On average, each audio lasts 10s and is annotated with 6 temporally-localized imperative sentences. 
We evaluate caption quality using METEOR score~\cite{banerjee2005meteor}. 
For an overall evaluation at time sense, we utilize event-based measures (Eb) and clip-level macro F1 score (At)~\cite{mesaros2016metrics} to evaluate the event localization performance and the diversity of descriptions between the generated events and the ground truth. We use a small language model to map each audio caption into its respective semantic category.

\textbf{For temporal audio grounding}, we use the AudioGrounding test data. The dataset contains 4662 audios and involves  10548 queries, where 9,551 pairs are used for training and 997 for testing. 
The average duration of the audios is 9.98 seconds and each audio contains 2.21 annotated pieces.
To evaluate temporal localization performance, we compute the Intersection over Union (IoU) between predicted and ground-truth timestamps. We then report the mean IoU (mIoU) as a measure of overall accuracy, and Recall at thresholds of {0.5, 0.7, 0.9}, which denotes the percentage of ground-truth events successfully retrieved by the model.

\textbf{For timeline speech summarization}, We evaluate on a subset of the CNN/DailyMail test set containing over 886 speech samples from articles under 1600 characters. We report standard content metrics (ROUGE~\cite{lin2004rouge} and METEOR) and a mean Intersection over Union (mIoU) score to measure how accurately the summary is grounded to the audio timeline.

\subsection{Appendix D: More Qualitative Results}
To illustrate the model's performance, we provide an extended set of qualitative results in Figures~\ref{fig:dac}-\ref{fig:tss}. These examples, which cover dense audio captioning, temporal audio grounding, and timeline speech summarization, highlight our model's ability to handle a diverse range of complex temporal localization tasks.

\end{document}